\documentclass[12pt]{iopart}  

\usepackage{graphicx}
\usepackage{dcolumn}
\usepackage{bm}

\newcommand{\physt}{${\cal T}$}

\def\agt{ \lower .75ex \hbox{$\sim$} \llap{\raise .27ex \hbox{$>$}} }
\def\alt{ \lower .75ex \hbox{$\sim$} \llap{\raise .27ex \hbox{$<$}} }

\begin{document}
	
\title[Multiverse conundrums and coincidences]{On making predictions in a multiverse: conundrums, dangers, and coincidences}

\author{Anthony Aguirre} 

\address{Department of Physics, UC Santa Cruz, Santa Cruz, CA 95064}
\ead{aguirre@scipp.ucsc.edu} 

\date{\today}

\begin{abstract}
The notion that there are many ``universes'' with different properties
is one answer to the question of ``why is the universe so hospitable
to life?'' This notion also naturally follows from current ideas in
eternal inflation and string/M theory. But how do we test such a
``multiverse'' theory: which of the many universes do we compare to
ours? This paper enumerates would would seem to be essential
ingredients for making testable predictions, outlines different
strategies one might take within this framework, then discusses some
of the difficulties and dangers inherent in these approaches.
Finally, I address the issue of whether there may be some {\em
general, qualitative} predictions that multiverse theories might
share.
\end{abstract}

\setcounter{footnote}{0}

\section{Introduction}

The standard model of particle physics and the standard model of
cosmology are both rife with numerical parameters that must have
values fixed by hand to explain the observed world.  The world would
be a radically different place if some of these constants took a
different value.  In particular, it has been
argued that if any one of six (or perhaps a few more) numbers did not
have rather particular values, then life as we
know it would not be possible~\cite{rees:2000}: atoms would not exist,
or no gravitationally bound structures would form in the universe, or some other
calamity would occur that would appear to make the (alter)-universe a
very dull and lifeless place.  How, then, did we get so lucky as to be
here?

This question is an interesting one because all of
the possible answers to it that I have encountered or devised entail
very interesting conclusions.  An essentially exhaustive list of such
answers is:
\begin{enumerate}
\item We just got very lucky: all of the numbers could have been very
different, in which case the universe would have been barren -- but
they just happened by pure chance to take values in the tiny part of
parameter space that would allow life. We owe our existence to one
very, very, very lucky roll of the dice.\footnote{If the parameters
were explained -- by some deeper theory -- in terms of fewer or no
parameters, this does not change much: it would explain the origin of
the parameters, but their hospitality to life would still be dumb
luck.}
\item We weren't particularly lucky: almost any set of parameters
would have been fine, because life would find a way to arise in nearly
any type of universe.  This is quite interesting because
it implies (at least theoretically) the existence of life forms
radically different from our own, existing for example in universes
with no atoms or with no bound structure, or overrun with black holes,
etc.
\item The universe was specifically designed for life. The choice of
constants only happened once, but their values were determined in some
way by the need for us to arise.  This might be divine agency, or some
radical form of Wheeler's ``self-creating universe", or super-advanced
beings that travel back in time to set the constants at the beginning
of the universe, etc.  However the reader feels about this
possibility, they must admit that it would be interesting if true.
\item We did not have to get lucky, because there are many universes
with different sets of constants -- i.e., the dice were rolled many,
many times.  We are necessarily in one of the universes that allows
life, just as we necessarily reside on a planet that supports life,
even when most others may not.  This is interesting because it means
that there are other very different universes coexisting with ours in
a ``multiverse".

\end{enumerate}

These four answers -- luck, {\em elan vital}, design, and multiverse
-- will appeal at different levels to different readers.  But I think
it is hard to argue that the multiverse is necessarily less reasonable
than the alternatives.  Moreover, as is discussed at length elsewhere
in this volume, there are quite independent reasons to believe, on the
basis of inflation, quantum cosmology, and string/M theory, that there
might quite naturally be many regions larger than our observable
universe, governed by different sets of low-energy physics.  I am not
aware of any independent scientific argument for the other three
possible explanations.

Whether they are contemplated as an answer to the ``why are we lucky"
question, or because they are forced upon us from other
considerations, multiverses come at a high price.  Even if we have in
hand a physical theory and cosmological model that lead to a
multiverse, how do we test it? If there are many sets of constants,
which ones do we compare to those we observe?  In the next section of
this chapter I will outline what I think a sound prediction in a
multiverse would look like.  As will become clear, this requires many
ingredients, and there are some quite serious difficulties in
generating some of these ingredients, even with a full theory in hand.
For this reason, many short-cuts have been devised to try to make
predictions more easily.  In the third section I will describe a
number of these, and show the cost that this convenience entails.
Finally, in Section~\ref{sec-coinc} I will focus on the interesting
question of whether the anthropic approach to cosmology might lead to
any {\em general} conclusions about how the study of cosmology will
look in coming years.

\section{Making predictions in a multiverse}
\label{sec-pred}

Imagine that we have a candidate physical theory and set of
cosmological boundary conditions (hereafter denoted \physt) that
predicts an ensemble of physically realized systems, each of which is
approximately homogeneous in some coordinates and can be characterized
by a set of parameters (i.e. the constants appearing in the standard
models of particle physics and cosmology; I assume here that the laws
of physics themselves retain the same form). Let us denote each such
system a ``universe'' and the ensemble a ``multiverse''.  Given that
we can observe only one of these universes, what conclusions can we
draw regarding the correctness of \physt, and how?

One possibility would be if there were a parameter for which {\em
none} of the universes in the ensemble had the value we observe.  In
this case \physt\ would be ruled out. (Note that any \physt\ in which
at least one parameter has a range of values that it does not take in
{\em any} universe is thus rigorously falsifiable, which is a nice thing
for a theory to be).  Or perhaps some parameter takes only one value
in all universes, and this value matches the observed one.  This would
obviously be a significant accomplishment of the theory. Both
possibilities are good as far as they go, and seem completely
uncontroversial. But they do not go far enough. What if our
observed parameter values appear in some but not all of the universes?
Could we still rule out the theory if those values are incredibly
rare, or gain confidence if they are extremely common?

I find it hard to see why not.  If some theory predicts outcome A of
some experiment with $p=0.99999999$ probability, and outcome B with
probability $1-p$, I think we would be reluctant to accept the theory
if a single experiment were performed and showed outcome B, {\em
even if we did not get to repeat the experiment}.  In fact, it seems
consistent with all normal scientific methodology to rule out the
theory at $99.999999\%$ confidence -- the problem is just that without
repeating our measurements we will not be able to {\em increase} this
confidence.  This seems to be exactly analogous to the multiverse {\em
if} we can compute, given our \physt, the {\em probability} that we should
observe a given value for some observable.

Can we compute this probability distribution in a multiverse?
Perhaps.  I will argue that to do so in a sensible way, we would need
seven successive ingredients.

\begin{enumerate}

\item First, of course, we require a multiverse: an ensemble of regions,
each of which would be considered a universe to observers inside it
(i.e. its properties would be uniform for as far as those observers
could see), but each of which may have different properties.  

\item Next we need to isolate the set of parameters characterizing the
different universes.  This might be the set of 20-odd free parameters
in the standard model of particle physics (see, e.g.,
Ref.~\cite{Hogan:1999wh} and references therein), plus a dozen or so
cosmological parameters~\cite{Aguirre:2001,Tegmark:2004qd}.  There
might be additional parameters that become important in other
universes, or differences (such as different forms of the physical
laws) that cannot be characterized by differences in a finite set of
parameters.  But for simplicity let us assume that some set of $N$
numbers $\alpha_i$ (where $i=1..N$) fully specify each universe.

\item Given our parameters, we need some {\em measure} with which to
calculate the multi-dimensional probability distribution $P(\alpha_i)$
for the parameters. We might, for example, ``count each universe
equally" to obtain the probability $P_U(\alpha_i)$, defined to be the
chance that a randomly chosen universe from the ensemble would have
the parameter values $\alpha_i$.\footnote{Note that this is really
shorthand for ${d{\cal P}\over
d\alpha_1..d\alpha_N}d\alpha_1..d\alpha_N$, the probability that
$\alpha_i$ are all within the interval
$[\alpha_i,\alpha_i+d\alpha_i]$, where ${\cal P}(\alpha_i)$ is a
cumulative probability distribution.}  This can be a bit tricky,
however, because it depends on how we delineate the universes: suppose
that $\alpha_1=a$ universes happen to be $10^{10}$ times larger than
$\alpha_1=b$ universes.  What would then prevent us from ``splitting"
each $\alpha_1=a$ universe into 10, or 100, or $10^{10}$ universes,
thus radically changing the relative probability of $\alpha_1=a$
vs. $\alpha_1=b$?  These considerations might lead us to take a
different measure such as volume, e.g. to define $P_V(\alpha_i)$, the
chance that a randomly point in space would reside in a universe with
parameter values $\alpha_i$.  But in an expanding universe volume
increases, so this would depend on the time at which we choose to
evaluate the volume in each universe.  We might then consider some
``counting" object that endures, say a baryon (which is relatively
stable), and define $P_B(\alpha_i)$, the chance that a randomly chosen
baryon would reside in a universe with parameter values
$\alpha_i$. But now we have excluded from consideration universes with
no baryons.  Do we want to do that?  This will be addressed in step
(v).  For now, note only that it is not entirely clear, even in
principle, which measure we should place over our multiverse.  We can
call this the ``measure problem."

\item Once we choose a measure object $M$, we still need to actually
compute $P_M(\alpha_i)$, and this may be far from easy.  For example,
in computing $P_V$, some universes may have infinite volume.  In this
case values of $\alpha_i$ leading to universes with finite volume will
have zero probability. How, though, do we compare two infinite volumes? The
difficulty can be seen by considering how we would count the fraction
of red vs. blue marbles in an infinite box.  We could pick one red,
then one blue, and find a 50-50 split.  But we could also repeatedly
pick 1 red, then 2 blue, or 5 red, then 1 blue.  We could do this
forever and so obtain any ratio we like.  What we would like to do is
just ``grab a bunch of marbles at random" and count the ratio.  But in
the multiverse case it is not so clear how to perform this random
ordering of marbles to pick. This difficulty, which might be termed
the ``ordering problem"~\cite{Tegmark:2004qd}, has been discussed a
number of times in the context of eternal
inflation~\cite{Linde:1995uf,Vanchurin:1999iv,Guth:2000ka,Garriga:2001ri}
and a number of plausible prescriptions have been proposed.  But there
does not seem to be any generic solution, or convincing way to prove that
one method is correct.

\item If we have managed to calculate $P_M(\alpha_i)$, do we have a
prediction?  Sortof.  We have an answer to the question: ``given that
I am (or can associate myself with) a randomly chosen $M$-object,
which sort of universe am I in?"  But this is {\em not} necessarily
the same as the more general question: ``what sort of universe am I
in?"  First of all,  different $M$-objects will generally give
different probabilities, and they cannot all be the answer to the same
question.  Second, we may not be all that closely associated with our
$M$-object (which was chosen mainly to provide {\em some} way to
compute probabilities) because it does not take into account important
requirements for our existence. For example, if $M$ were volume, I
would be asking what I should observe given that I am at a random
point in space; but we are {\em not} at a random point in space (which
would on average have a density of $10^{-29}\,$g/cc), but rather at one
of the very rare points with density $\sim 1$\,g/cc.  The reason for this
improbable situation is obviously ``anthropic" -- we just do not worry
about it because we can observe many other regions at the proper
density (if we could not see such regions, we might be more reluctant
to accept a cosmological model with such a low average density.)
Finally, it might be argued that the question we have answered through
our calculation is not nearly as specific a question as we could ask,
because we know a lot more about the universe than that it contains
volume, or baryons.  We might, instead, ask ``given that I am in a
universe with the properties we have already observed, what should I
observe in the future?"

\ \ \ \ As discussed at length in~\cite{Aguirre:2004qb}, these
different specific questions can be usefully thought of as arising
from different choices of conditionalization.  The probabilities
$P_M(\alpha_i)$ are conditioned on as little as possible, whereas the
anthropic question of ``given that I am a randomly chosen {\em
observer}, which should I measure" specifies probabilities conditioned
on the existence of an ``observer", while the approach of ``given what
I know now, what will I see" specifies probabilities conditioned on
being in a universe with all of the properties that we have already
observed.  These are three genuinely different approaches to making
predictions in a multiverse that may be termed, respectively,
``bottom-up", ``anthropic", and ``top down".

\ \ \ \ Let us denote by $O$ the conditionalization object used to
specify these conditional probabilities.  In bottom-up reasoning, it
would be the same as the $M$-object; in the anthropic approach it
would be an ``observer", and in the top-down approach it could be a
universe with the currently-known properties of our universe.  It can
be seen that they inhabit a spectrum, from the weakest
conditionalization (bottom-up) to the most stringent (top-down).  Like
our initial $M$-object, choosing a conditionalization is unavoidable
and important, and there is no obviously correct choice to make. (See
Refs.~\cite{bostrum,Hartle:2004qv} for similar conditionalization
schemas.)

\item Having decided on a conditionalization object $O$, the next step
is to compute the number of $N_{O,M}(\alpha_i)$ of $O$-objects per
$M$-object, for each set of values of the parameters $\alpha_i$.  For
example, if we have chosen to condition on observers, but have used
baryons to define our probabilities, then we need to calculate the
number of observers per baryon as a function of cosmological
parameters.  We can then calculate $P_O(\alpha_i)\equiv
P_M(\alpha_i)N_{O,M}(\alpha_i)$, i.e. the probability that a randomly
chosen $O$-object (observer) resides in a universe with parameters
$\alpha_i$.  There are a few possible pitfalls in doing this.  First,
if $N_{O,M}$ is infinite, then the procedure clearly breaks because
$P_O$ then becomes undefined.  This is why the $M$-object should be
chosen to requires as little as possible for its existence (and hence
be associated with the minimal-conditionalization bottom-up approach).
This difficulty will generically occur if the existence of an
$O$-object does not necessarily entail the existence of an
$M$-object. For example, if the $M$-object were a baryon but the
$O$-object were a bit of volume, then $N$ would be infinite for
$\alpha_i$ corresponding to universes with no baryons.  The problem
arises because baryons require volume to be in, but volume does not
require a baryon to be in it.  This seems straightforward, but gets
much murkier when we consider the {\em second} difficulty when
calculating $N$, which is that we may not be able to precisely define
what an $O$-object is, or what it takes to make one.  If we say that
the $O$-object is an observer, what exactly does that mean?  A human?
A carbon-based life form? Can observers exist without water? Without
heavy elements?  Without baryons?  Without volume?  It seems quite
hard to say.  We are forced, then, to choose some proxy for an
observer, e.g. a galaxy, or a star with possible planets, etc.  But
our probabilities will perforce depend on the chosen proxy and this
must be kept in mind.

\ \ \ \ It is worth noting a small bit of good news here. If we do
manage to consistently compute $N_{O,M_1}$ for some measure object
$M_1$, then insofar as we want to condition our probabilities on
$O$-objects, we have solved the measure problem: if we could
consistently calculate $N_{O,M_2}$ for a different measure object
$M_2$, then we should obtain the same result for $N_O$,
i.e. $N_{O,M_1}P_{M_1}=N_{O,M_2}P_{M_2}$. Thus our choice of $M_1$
(rather than $M_2$) becomes unimportant.

\item The final step in making predictions is to make the assumption
that the probability that we will measure some set of $\alpha_i$ is
given by the probability that a randomly chosen $O$-object will.  This
assumption really entails two others: first, that we are some how
directly associated with $O$-objects, and second that we have not,
simply by bad luck, observed highly improbable values of the
parameters. The assumption that we are {\em typical} observers has
been termed the ``principle of mediocrity"~\cite{vil:95}.  One may
argue about this assumption, but {\em some} assumption is necessary if
we are to connect our computed probabilities to observations, and it
is difficult to see what alternative assumption would be more
reasonable.

\end{enumerate}

The result of all this work would be the probability $P_O(\alpha_i)$
that a randomly selected $O$-object (out of all of the $O$-objects
that exist in multiverse) would reside in a universe governed by
parameters $\alpha_i$, along with a reason to believe that this same
probability distribution should govern what we will observe.  We can
then make the observations (or consider some already-made ones). If
the observations are highly improbable according to our predictions,
we can rule out the candidate \physt\ at some confidence that depends
on how improbable our observations were.  Apart from the manifest and
grave difficulties involved in actually completing the seven listed
steps in a convincing way, I think the only real criticism that can be
leveled at this approach is that unless $P=0$ for our observed
paramters, there will always be the chance that the \physt\ was
correct and we measured an unlikely result.  Usually, we can rid
ourselves of this problem by repeating our experiments to make $P$ as
small as we like (at least in principle), while here we do not have
that option -- once we have ``used up" the measurement of all of the
paramters required to describe our universe (which appears to be
rather surprisingly few, at least according to current theories), we
are done.

\section{Making predictions in a multiverse more easily: a bestiary of shortcuts}
\label{sec-anthrop}

Although the idea of a multiverse has been around for quite a while,
no one has ever really come close to making the sort of calculation
outlined in the previous section.\footnote{The most ambitious attempt
is probably the recent one by Tegmark~\cite{Tegmark:2004qd}.} Instead,
those wishing to make predictions in a multiverse context have made
strong assumptions about which parameters $\alpha_i$ actually vary
across the ensemble, about the choice of $O$-object, and about the
quantities $P_M(\alpha_i)$ and $N_{O,M}(\alpha_i)$ that go into
predicting their probabilities for measurement.  Some of these
shortcuts aim simply to make a calculation tractable; others are
efforts to avoid anthropic considerations, or alternatively to use
anthropic considerations to avoid other difficulties.

I would not have listed any ingredients that I thought could be
omitted from a really sound calculation, thus all of these shortcuts
are necessarily incomplete (some, in my opinion, disastrously so).
But by listing and discussing them, I hope to give the reader both a
flavor for what sort of anthropic (or anthropic-esque) arguments have
been made in the literature, and where they may potentially go astray.

\subsection{The ``Maybe anthropic considerations only allow one set of parameters" hope}

This assumption underlies a sort of anthropic reasoning that has
earned the anthropic principle a lot of ill will.  It goes something
like: ``Let's assume that lots of universes governed by lots of
different parameter values exist.  Then since only universes with
parameter values almost exactly the same as ours allow life, we must
be in one of those, and we should not find it strange if our parameter
values seem special.''  In the conventions I have described, this is
essentially equivalent setting $O$-objects to be observers, then
hoping that the ``hospitality factor" $N_{\rm obs,M}(\alpha_i)$ is
very narrowly peaked around one particular set of parameters.  In this
case, the {\em a priori} probabilities $P_M$ are pretty much
irrelevant because the shape of $N_{\rm obs,M}$ will pick out just one
set of parameters.  Because our observed values $\alpha_i^{\rm obs}$
definitely allow observers, the allowed set must then be very near
$\alpha_i^{\rm obs}$.

Three problems with this type of reasoning are as follows.  First, it
is rather circular: it entails picking the $O$-object to be an
observer, but then quickly substituting a ``universe just like ours"
for the $O$-object, with the reasoning that such universes will
definitely support life.\footnote{One can also argue that if there
were other, more common, universes that supported life, we ought to be
in them; since we are, not, we should assume that almost all
life-supporting universes are like ours. But this is also circular in
assuming that the whole anthropic argument works, in formulating the
argument.}  Thus we have arrived at: the universe we observe should be
pretty much like the observed universe.  The way to avoid this
silliness is to allow at least the possibility that there are
life-supporting universes with $\alpha_i \neq \alpha_i^{\rm obs}$,
i.e. to discard the unproven {\em assumption} that $N_{\rm obs,M}$ has
a single, dominant, narrow peak.

The second problem is that if $N_{\rm obs,M}$ were really {\em so} narrowly
peaked as to render $P_M$ irrelevant, then we would be in serious
trouble as theorists, because we would lose any ability to distinguish
between candidates for our fundamental theory: unless our observed
universe is {\em impossible} in the theory, then the anthropic factor
would force the predictions of the theory to match our
observations. As discussed in the next section, this is not good.

The third problem with $N_{\rm obs,M}$ being an extremely peaked function is
that it does not appear to be true! As discussed in below, it
appears that for any reasonable surrogate for observers (e.g. galaxies
like ours, or stars with heavy elements, etc.), calculations done
using our current understanding of galaxy and structure formation
indicate that the region of parameter space in which there can be many
of those objects may be small compared to the full parameter space, but it
is much larger than the region compatible with our observations.

\subsection{The ``Just look for zero probability regions ($P=0$) in parameter space" approach}

As mentioned in Section 2, there is a (relatively!) easy thing to do
with a multiverse theory \physt: work out which parameters
combinations cannot occur in {\em any} universe.  If the combination
we actually observe is one of these, then the theory is ruled out.
This is unobjectionable, but a rather weak way to test a theory
because given two theories that are {\em not} ruled out, we have no
way whatsoever of judging one to be better, even if the parameter
values we observe are in some sense generic in one and absurdly rare
in the other.\footnote{Amusingly, in terms of testing \physt, this
approach which makes {\em no} assumptions about $N_{O,M}$ is
equivalent to the approach just described of making the very strong
assumption that $N_{O,M}$ allows only one specific set of parameter
values, because in either case a theory can only be ruled out if our observed values are impossible in that theory.}

This is not how science usually works.  For example, suppose our
theory is that a certain coin-tossing process is unbiased.  If our only way to test this theory was to look for experimental outcomes that are impossible, then
the theory would unfalsifiable: we would have to
accept it theory for {\em any} coin we are confronted with, because {\em
no} sequence of tosses would be impossible in it! Even if 10,000
tosses in a row all came up heads, we would have no grounds for
doubting our theory because while getting heads 10,000 times in a row on a fair coin is absurdly improbably, it is not impossible. Nor would we have reason to prefer the
(seemingly much better) ``nearly every toss comes out heads"
theory. Clearly this is a situation we would like to improve on, as
much in universes as in coin tosses.

\subsection{The ``Let's look for overwhelmingly more probable values" suggestion}

One possible improvement would be to assume that we will observe a
``typical" set of parameters in the ensemble, i.e. that we will employ
``bottom-up" reasoning as described in the first section, by using the
{\em a priori} (or ``prior") probabilities $P_M$ for some choice of
measure-object such as universes, and just ignore the
conditionalization factor $N_{O,M}$.  There are two possible
justifications for this. First, we might simply want to avoid any sort
of anthropic issues on principle.  Second, we might hope that some
parameter values are much, much more common than others, to the extent
that the $N_{O,M}$-factor becomes irrelevant -- in other words that
$P_M$ (rather than $N_{O,M}$) is a very strongly peaked around some
particular parameters.

The problem with this approach is the ``measure problem" discussed
above: there is an implicit choice of basing probabilities on
universes (say) rather than on (say) volume elements or baryons.  Each
of these measures has problems -- for example, it seems that
probabilities based on ``universes" depends on how the universes are
delineated, which can be ambiguous. Moreover there seems to be no
reason to believe that predictions made using any two measures should
agree particularly well.  For example, as discussed elsewhere in this
volume, in the string theory ``landcape" there are many possible
parameter sets, depending on which metastable minimum one chooses in a
potential that depends in turn on a number of fluxes that can take a
large range of discrete values.  Imagine that exponentially many more
minima lead to $\alpha_1=a$ than lead to $\alpha_1=b$.  Should we
expect to observe $\alpha_1=a$?  Not necessarily, because the relative
number of $a$-universes vs. $b$-universes that {\em actually come into
existence} may easily differ exponentially from the relative number of
$a$-minima vs. $b$-minima. (This seems likely to me in an
eternal-inflation context, where the relative number universes could
depend on exponentially-suppressed tunnelings between vacua.) Worse
yet, these may in turn differ exponentially (or even by an infinite factor)
from the relative numbers of baryons, or relative volumes.

In short, while we are free to use bottom-up reasoning with any choice
of measure object we like, we are not free to assert that other
choices would give similar predictions, or that conditionalization can
be rendered irrelevant.  So we had better have a pretty good reason
for the choice we make.

\subsection{The ``Let's fix some parameters to the observed values and predict others" shortcut}

Another way in which one might hope to circumvent anthropic issues is
to condition the probabilities on some or all observations that have
already been made.  In this ``top-down" (or perhaps ``pragmatic")
approach we ask: given everything that has been observed so far, what
will we observe in some future measurement?  It has a certain appeal,
as this is often what is done in experimental science: we do not try
to {\em predict} what our laboratory will look like, just what will
happen given that the lab is in a particular state at a given time.
In the conventions of Section 2, the approach could consist of
choosing the $O$-object to be universes with parameters agreeing with
the measured values.

While appealing, this approach suffers some deficiencies:
\begin{itemize}
\item It still does not completely avoid the measure problem, because
even once we have limited our consideration to universes that match
our current observations, we must still choose a measure with which to
calculate the probabilities for the remaining ones.

\item Through our conditioning, we may accept theories for which our
parameter values are wildly improbable, without supplying any
justification as to why we observe such improbable values.  This is
rather strange.  Imagine that I have a theory in which the
cosmological constant $\Lambda$ is (with very high probability) much
higher than we observe, and the dark matter particle mass $m_{\rm DM}$
is almost certainly $> 1000\,$GeV.  I condition on our observed
$\Lambda$, simply accepting that I am in an unusual universe.  Now say
I measure $m_{\rm DM}=1\,$GeV. I would like to say my theory is ruled
out.  Fine, but here is where it gets odd: according to top-down
reasoning, I should also have already ruled it out if I had done my
calculation in 1997, before $\Lambda$ was measured.  And someone who
invented the very same theory next week -- but had not been told that
I have already ruled it out -- would {\em not} rule it out, but
instead just take the low value of $m_{\rm DM}$ (along with the observed $\Lambda$) as part of the
conditionalization!

\item If we condition on everything we have observed, we obviously
give up the possibility of {\em explaining} anything we have observed
(which at this point is quite a lot in cosmology) through our theory.

\end{itemize}

The last two issues motivate variations on the top-down approach in which
only some current observations are conditioned on. Two of which I am aware are:

\begin{enumerate}

\item We might {\em start} by conditioning on all observations, then
progressively condition on less and less and try to ``predict" the
things we have decided not to condition on (as well, of course, as any
{\em new} observations)\cite{Hawking:2002af,Albrecht:2002uz}.  The
more we can predict, the better our theory is.  The problem is that
either (a) we will get to the point where we are conditioning on as
little as possible (the bottom-up approach), and hence the whole
conditionalization process will have been a waste of time, or (b) we
will still have to condition on some things, and admit either that
these have an anthropic explanation, or that we just choose to
condition on them (leading to the funny issues discussed above).

\item We might choose at the outset to condition on things that we
think may be fixed anthropically (without trying to actually generate
this explanation), then try to predict the others~\cite{Dine:2004ct}.
This is nice in being relatively easy, and in providing a
justification for the conditionalization.  It suffers from the
problems of (a) guessing which parameters are anthropically important
and which are not, (b) even if a parameter is anthropically
unimportant, it may be strongly correlated in $P_M$ with one that is,
and (c) we still have to face the measure problem, which we cannot
avoid by counting conditioning on observers, because we are avoiding
anthropic considerations.

\end{enumerate}

\subsection{The ``Let's assume just one parameter varies" simplification}

Most of the ``shortcuts" discussed so far have been attempts to avoid
anthropic considerations.  But we may, instead, consider how me might
try to formulate an anthropic prediction (or explanation) for some
observable, without going through the full calculation outlined in
Section 2.  The way of doing this that has been employed in the
literature (largely in the efforts of Vilenkin and collaborators) is as follows.

First, one fixes all but one (or perhaps two) of the parameters to the
observed values.  This is done for tractability and/or because one hopes
that they will have non-anthropic explanations. Let us call the
parameter that is allowed to vary across the ensemble $\alpha$.

Next, an $O$-object is chosen such that {\em given that only $\alpha$
varies}, it is hoped that (a) the number $N_{O,M}$ of these objects
(per baryon, or per comoving volume element) in a given universe is
calculable, and (b) this number is arguably proportional to the number
of observers.  For example, if only $\Lambda$ varies across the
ensemble, galaxies might make reasonable $O$-objects because a
moderately different $\Lambda$ will probably not change the number of
observers per galaxy, but {\em will} change the number of galaxies in
a way that can be computed using fairly well-understood theories of
galaxy and structure formation (for examples
see~\cite{Garriga:1999hu,Garriga:1999bf,Garriga:2000cv,Tegmark:2003ug,Pogosian:2004hd,Graesser:2004ng,Tegmark:2004qd}).

Third, it is assumed that $P_M(\alpha)$ is either flat or a simple
power-law, without any complicated structure.  This can be done just
for simplicity, but it is often argued to be
natural~\cite{Vilenkin:1995nb,Weinberg:2000qm,Smolin:2004yv}.  The
flavor of this argument is as follows. If $P_M$ is to have interesting
structure over the relatively small range in which observers are
abundant, there must be a parameter of order the observed $\alpha$ in
the expression for $P_M$.  But precisely this absence is what
motivated the anthropic approach.  For example, if the expression for
$P_M(\Lambda)$ contained the energy scale $\sim 0.01\,$eV
corresponding to the observed $\Lambda$, the origin of that energy
scale would probably be more interesting than our anthropic argument,
as it would provide the basis for a (non-anthropic) solution to the
cosmological constant problem!

Under these (fairly strong) assumptions we can then actually calculate
$P_O(\alpha)$ and see whether or not the observed value is reasonably
probable given this predicted distribution.  For example when
$\Lambda$ alone is varied, a randomly chosen galaxy is predicted to lie in a
universe with $\Lambda$ comparable to (but somewhat larger than) the
value we see~\cite{Garriga:1999bf}.\footnote{This is for a ``flat"
probability distribution $dP_M/d\lambda\propto \lambda^\alpha$ with
$\alpha=0$. For $\alpha > 0$, higher values would be predicted, and
with $\lambda < 0$ lower values would be favored.}

I actually think this sort of reasoning is pretty respectable, {\em
given the assumptions made}. In particular, the anthropic argument in
which {\em only} $\Lambda$ varies is a relatively clean one. But there
are a number of pitfalls when it is applied to parameters other than
$\Lambda$, or when one allows multiple parameters to vary
simultaneously.

\begin{itemize}

\item Assuming that the abundance of observers is strictly
proportional to that of galaxies only makes sense if the number of
galaxies -- and not their properties -- changes as $\alpha$
varies. However, changing nearly any cosmological parameter will
change the properties of typical galaxies.  For example, increasing
$\Lambda$ will decrease galaxy numbers, but also make galaxies smaller
on average, because a high $\Lambda$ squelches structure formation at
late times when massive galaxies form.  Increasing the amplitude of
primordial perturbations would similarly lead to smaller, denser --
but more numerous -- galaxies, as would increasing ratio of dark
matter to baryons.  In these cases, we must specify in more detail
what properties an observer-supporting galaxy should have, and this is
very difficult to do without falling into the circular-argument trap
of assuming that only galaxies like ours support life.  Finally, this
sort of strategy seems unlikely to work if we try to change {\em
non}-cosmological parameters, as this could lead to radically
different physics and the necessity of thinking {\em very} hard about
what sort of observers there might be.

\item The predicted probability distribution clearly depends on $P_M$,
and the assumption that $P_M$ is flat, or a simple power law, can
break down.  This can happen even for $\Lambda$~\cite{Garriga:1999bf}
but perhaps more naturally for other parameters such as the dark
matter density for which particle physics models can already yield
sensible values. Moreover, this breakdown is much more probable if (as
discussed below and contrary to the assumption made above) the
hospitality factor $N_{O,M}(\alpha)$ is significant over many orders
of magnitude in $\alpha$.

\item Calculations of the hospitality factor $N_{O,M}(\alpha)$ can go
awry if $\alpha$ is changed more than a little.  For example, a
neutrino mass slightly larger than we observe would suppress galaxy
formation by erasing small-scale structure.  But neutrinos with a
large ($\agt 100\,$eV) mass would act as dark matter and lead to
strong halo formation.  Whether these galaxies would be hospitable is
questionable (they would be very baryon-poor), but the point is that
the physics becomes {\em qualitatively} different. As another example,
a lower photon/baryon ratio $n_\gamma/n_b$ would lead to
earlier-forming, denser galaxies.  But a {\em much} smaller value
would lead to qualitatively different structure formation, as well as
the primordial generation of heavy elements~\cite{Aguirre:2001}.  As
discussed at length in ref.~\cite{Aguirre:2001}, these changes are
very dangerous because over orders of magnitude in $\alpha$,
$P_M(\alpha)$ will tend to change by many orders of magnitude. Thus
even if these alter-universes only have a few observers in them, they
may dominate $P_O$ and hence qualitatively change the predictions.

\item Along the same lines, but perhaps even more pernicious, when
multiple parameters are varied simultaneously, the effects of some
variations can offset the effect of others so that universes quite
different from ours can support many of our chosen $O$-objects.  For
example, increasing $\Lambda$ cuts off galaxy formation at a given
cosmic density, but raising the perturbation amplitude $Q$ causes
galaxies to form earlier (thus nullifying the effect of $\Lambda$).
This can be seen in the calculations
of~\cite{Tegmark:1997in,Garriga:2000cv}, and is discussed explicitly
in ~\cite{Aguirre:2001,Graesser:2004ng,Tegmark:2004qd}.  Many such
deneneracies exist, because rasing $\Lambda$, $n_\gamma/n_b$, or the
neutrino mass all decrease the efficiency of structure formation,
while raising $\Omega_{\rm DM}/\Omega_b$ or $Q$ increase the
efficiency.  As an extreme case, it was shown in~\cite{Aguirre:2001}
that if $Q$ and $n_\gamma/n_b$ are allowed to vary with $\Lambda$,
then universes with $\Lambda$ of $10^{17}$ times our observed value
could arguably support observers!  Including more cosmological
parameters, or non-cosmological parameters, can only make this problem
worse.

\end{itemize}

These problems indicate that while anthropic arguments concerning
$\Lambda$ in the literature are relatively ``clean", it is unclear
whether other parameters (taken individually) will work as nicely.
More importantly, a number of issues arise when several parameters are
allowed to vary at once, and there does not seem to be any reason to
believe that success in explaining one parameter anthropically will
persist when additional parameters are allowed to vary.  In some
cases, it may: for example, allowing neutrino masses to vary in
addition to $\Lambda$ does not appear to spoil the anthropic
explanation of a small but nonzero cosmological
constant~\cite{Pogosian:2004hd}.  On the other hand, allowing $Q$ to
vary does, unless $P_M(Q)$ is strongly peaked at small values of
$Q$~\cite{Graesser:2004ng}.  I suspect that allowing $\Omega_{\rm
DM}/\Omega_b$ or $n_\gamma/n_b$ to vary along with $\Lambda$ would
have a similar effect.

\subsection{So what should we do?}

For those serious about making predictions in a multiverse, I would
propose that rather than working to generate additional incomplete
anthropic arguments by taking shortcuts, a much better job must be
done in each of the individual ingredients.  For one example, our
understanding of galaxy formation is sufficiently strong that the
multi-dimensional hospitality factor $N_{O,M}(\alpha_i)$ could
probably be computed for $\alpha_i$ within a few orders of magnitude
of the observed values, for $O-$objects of galaxies with properties
within some range. Second, despite some nice previous work, I think
the problem of how to compute $P_M$ in eternal inflation is a pretty
open one. Finally, the string/M theory landscape (which is generating
a lot of interest in the present topic right now) cannot hope to say
much of anything about $P_M$ until its place in cosmology is
understood -- in particular, we need both a better understanding of
the statistical distribution of field values that result from
evolution in a given potential, and also an understanding of how
transitions between vacua with different flux values occur, and
exactly what is transitioning.

\section{Cosmic coincidences and living dangerously: are there general predictions of anthropic reasoning?}
\label{sec-coinc}

The preceding sections should have suggested to the reader that it
will be a huge project to compute a sound prediction of cosmological
and physical parameters from a multiverse theory in which they vary.
It may be so hard that it will be a very long time before any such
calculation is at all believable.  It is worth asking then: is there
any way nature might give us an indication as to whether the anthropic
approach is a sensible one, i.e. does the anthropic approach make any
sort of {\em general} predictions even without the full calculation of
$P_O$? Interestingly, I think the answer might be yes: I am aware of
two such general (though somewhat vague) predictions of the anthropic
approach.

To understand the first, assume that only one parameter, $\alpha$,
varies, and consider $p(\log\alpha) = \alpha P_M(\alpha)$, the
probability distribution in $\log\alpha$, given by some theory \physt.
For $\log\alpha$ near the observed value $\log\alpha^{\rm obs}$, $p$
can basically only be doing one of three things: it can rise with
$\log \alpha$, fall with $\log \alpha$, or be approximately constant.
In the first two cases, the theory \physt\ would predict that we
should see a value of $\alpha$ that is, respectively, higher or lower
than we actually do {\em if} no anthropic conditionalization $N_{\rm
obs, M}$ is applied.  Now suppose we somehow compute $N_{\rm obs,
M}(\alpha)$ and find that it falls off quickly for values of $\alpha$
much smaller or larger than we observe, i.e. that only a range
$\alpha_{\rm min} \alt \alpha \alt \alpha_{\rm max}$ is
``anthropically acceptable".  Then we have an anthropic argument
explaining $\alpha^{\rm obs}$, because this falloff means that $P_{\rm
obs}$ will only be significant near $\alpha^{\rm obs}$.  But now note
that {\em within} the anthropically acceptable range, $P_{\rm obs}$
will be peaked near $\alpha_{\rm max}$ if $p$ is increasing with
$\alpha$, or near $\alpha_{\rm min}$ if $p$ is decreasing with
$\alpha$.  That is, we should expect $\alpha^{\rm obs}$ at one edge of
the anthropically acceptable range.  This idea has been called the
``principle of living dangerously"~\cite{Dimopoulos:2003iy}.  It
asserts that for a parameter that is anthropically determined, we
should expect that a calculation of $N_{\rm obs,M}$ would reveal that
observers would be strongly suppressed either for $\alpha$ slightly
larger or slightly smaller than $\alpha_{\rm obs}$, depending on
whether $p$ is rising or falling.

Now, this is not a very specific prediction: exactly where we would
expect $\alpha^{\rm obs}$ to lie depends both on how steep
$p(\log\alpha)$ is, and how sharp the cutoff in $N_{\rm obs,M}$ is for
$\alpha$ outside of the anthropically acceptable range.  And it would
not apply to anthropically-determined parameters in all possible
cases.  (For example, if $p$ were flat near $\alpha^{\rm obs}$, but
also very high at $\alpha \gg \alpha^{\rm obs}$, anthropic effects
would be required to explain why we do not observe the very high
value; but any region within the anthropically acceptable range would
be equally probable, so we would not expect to be, so to speak, living
on the edge.)  Despite these caveats, this is a prediction of sorts,
because the naive expectation would probably be for our observation to
place us somewhere in the interior of the region of parameter space
that is hospitable to life, rather than at the edge.

A second sort of general prediction of anthropic reasoning is
connected to what might be called ``cosmic coincidences."  For
example, many cosmologists have asked themselves (and each other) why
the current density in vacuum energy, dark matter, baryons, and
neutrinos are all within a couple of orders of magnitude of each other
-- making the universe a much more complicated place than it might be.
Conventionally, it has been assumed that these coincidences are just
that, and follow directly from fundamental physics that we do not
understand.  But if the anthropic approach to cosmology is really
correct (that is, if it is the real answer to the question of why
these densities take the particular values they do), then the
explanation is quite different: the densities are bound together by
the necessity of observers' existence, because only certain
combinations will do.

More explicitly, suppose several cosmological parameters are governed
by completely unrelated physics, so that their individual prior
probabilities $P_M$ simply multiply to yield the multidimensional
probability distribution. For example, we might have
$P_M(\Lambda,\Omega_{\rm DM}/\Omega_b,Q)=P_M(\Lambda)P_M(\Omega_{\rm
DM}/\Omega_b)P_M(Q)$.  But even if $P$ factors, the hospitality factor
$N_{\rm obs,M}$ will almost certainly not: if galaxies are
$O$-objects, the number of galaxies formed at a given $\Lambda$ will
depend on both other parameters, and only certain combinations will
give a significant number of observers.  Thus $P_O=N_{\rm obs,M}P_M$
will likewise have correlations between the different parameters that
lead to only particular combinations (for example those with
$\Omega_{\rm DM}/\Omega_b\sim 1-10$ for a given $Q$ and $\Lambda$)
having high probabilities.  The cosmic coincidences would be explained
in this way.

This anthropic explanation of coincidences, however, should not only
apply to things that we have already observed.  If it is correct, then
it should apply also to {\em future} observations; that is, we should
expect to uncover yet more bizarre coincidences between quantities
that seem to follow from quite unrelated physics.

How might this actually happen?  Consider dark matter.  We know fairly
precisely how much dark matter there is in the universe, and what its
basic properties are.  But we have no real idea what it actually is,
and there are many, many possible candidates that have been proposed
in the literature.  In fact, we have no {\em observational} reason to
believe that dark matter is one substance at all: in principle it
could be equal parts axions, supersymmetric particles, and primordial
black holes.  The reason most cosmologists do not expect this
is that it would be a strange coincidence if three substances
involving quite independent physics all wound up with essentially the
same density in our universe.  But of course this would be just like the 
suprising-but-true coincidences that hold in already-observed
cosmology.

In the anthropic approach, these comparable densities could be quite
natural~\cite{Aguirre:2004qb}.  To see why, imagine that there are two
{\em completely independent} types of dark matter permeating the
ensemble: in each universe, they have some particular densities
$\rho_1$ and $\rho_2$ out of a wide range of possibilities, so that
the densities in a randomly chosen universe (or around a randomly
chosen baryon, etc.) will be given probabilistically by
$P_M(\rho_1)P_M(\rho_2)$.  Under these assumptions there is no reason
to expect that we should observe $\rho_1\sim\rho_2$ based just on
these {\em a priori} probabilities.  Now suppose, though, that
$N_{O,M}$ picks out a particular narrow range of {\em total} dark
matter density as anthropically acceptable.  That is,
$N_{O,M}(\rho_1+\rho_2)$ is narrowly peaked about some $\rho_{\rm
anth}$. In this case, the peak of the probability distribution
$P_O(\rho_1,\rho_2)$, which indicates what values a randomly chosen
observer should see, will occur where $P_M(\rho_1)P_M(\rho_2)$ is
maximized {\em subject to the condition} that $\rho_1+\rho_2\simeq
\rho_{\rm anth}$.  For simplicity let both prior probabilities be
power laws: $P_M(\rho_1)\propto \rho_1^\alpha$ and $P_M(\rho_2)\propto
\rho_2^\beta$. Now the coincidence: it is not hard to show that if
$\alpha \ge 0$ and $\beta \ge 0$, then the maximum will occur when
$\rho_1/\rho_2=\alpha/\beta$.  That is, the two components are likely
to have similar densities unless the {\em power law indices} of their
probability distributions differ by orders of
magnitude.\footnote{Extremely high power law indices are uncomfortable
in the anthropic approach because they would lead to $P_O$ being
peaked where $N_{O,M}$ is declining, i.e. we should be living {\em
outside} the anthropically comfortable range, not just dangerously but
downright recklessly.} Of course, there are many ways in which this
coincidence could fail to occur (e.g. negative power-law indices, or
correlated probabilities), but the point is that there is a quite
natural set of circumstances in which the components are coincident,
even though the {\em fundamental} physics is completely unrelated.

\section{Conclusions}

The preceding sections should have convinced the reader that there are
good reasons for scientists to be very worried if we live in a
multiverse: in order to test a multiverse theory in a sound manner, we
must perform a fiendishly difficult calculation of $P_O(\alpha_i)$,
the probability that an $O$-object will reside in a universe
characterized by parameters $\alpha_i$. And because of the
shortcomings of the shortcuts one may (and presently must) take in
doing this, almost any particular multiverse prediction is going to be
easy to criticize; only a quite good calculation is going to be at all
convincing.  Much worse, we face an unavoidable and important choice
in what $O$ should be: a possible universe, or an existing universe,
or a universe matching current observations, or a bit of volume, or a
baryon, or a galaxy, or an ``observer'', etc. 

I find it disturbingly plausible that ``observers" really are the
correct conditionalization object, that their use as such is the
correct answer to the measure problem, and that anthropic effects are
the real explanation for the values of some parameters (just as for
the local density that we observe).  Many cosmologiest appear to
believe that taking the necessity of observers into account is shoddy
thinking, and is employed only because it is the {\em easy} way out of
solving problems the ``right" way.  But the arguments of this paper
suggest that the truth may well be exactly the opposite: the anthropic
approach may be the right thing to do in principle, but nearly
impossible in practice. 

Nonetheless, we cannot do away with multiverses just by wishing them
away: we may in fact live in one, whatever the inconvenience to
cosmologists. The productive strategy then seems to be one of
accepting multiverses as a possibility, and working toward
understanding how to calculate the various ingredients necessary to
make predictions in one. Whether really performing such a calculation
will turn out to be possible, but it is certainly impossible if not
attempted.

Even if we cannot calculate $P_O$ in the foreseeable future, however,
cosmology in a multiverse may not be completely devoid of predictive
power.  For example, of anthropic effects are at work, they should
leave certain clues. First, if we could determine the region of
parameter space hospitable to observers, we should find that we are
living in the outskirts of the livable region, rather than somewere in
its midst.  Second, if the anthropic effects are the explanation of
the parameter values -- and coincidences between them -- that we see,
then it ought to predict that new coincidences will be observed in
future observations.

If in the next several decades dark matter is resolved into several
equally important components, dark energy is found to be three
independent substances, and several other ``cosmic coincidences'' are
observed, even someone the most die-hard skeptics might accede that
the anthropic approach may have validity -- why else would be universe
be so very baroque?  On the other hand, if we are essentially finished
in defining the basic cosmological constituents, and the defining
parameters are in the midst of a relatively large region of parameter
space that might arguably support observers, then I think the
anthropic approach would lose almost all appeal it has; we would be
forced to ask: why isn't the universe much wierder?

\end{document}